\documentclass[final,5p,times,twocolumn]{elsarticle} 

\usepackage{hyperref}
\usepackage{subfigure}

\makeatletter
\renewcommand{\@thesubfigure}{(\alph{subfigure})}
\renewcommand{\p@subfigure}{Figure\space}
\renewcommand{\p@figure}{Figure\space}
\makeatother %

\journal{Nuclear Instruments and Methods A}

\begin{document}

\begin{frontmatter}

\title{Performance of Large Area Picosecond Photo-Detectors (LAPPD\textsuperscript{TM})}


\author[mymainaddress]{A. V.  Lyashenko\corref{mycorrespondingauthor}}
\cortext[mycorrespondingauthor]{Corresponding author}
\ead{alyashenko@incomusa.com}

\author[mymainaddress]{B. W. Adams}
\author[mymainaddress]{M. Aviles}
\author[mymainaddress]{T. Cremer}
\author[mymainaddress]{C. D. Ertley}
\author[mymainaddress]{M. R. Foley}
\author[mymainaddress]{M. J. Minot}
\author[mymainaddress]{M. A. Popecki}
\author[mymainaddress]{M. E. Stochaj}
\author[mymainaddress]{W. A. Worstell}
\author[anl]{J. W. Elam}
\author[anl]{A. U. Mane}
\author[ssl]{O. H. W. Siegmund}
\author[uc]{H. J. Frisch}
\author[uc]{A. L. Elagin}
\author[uc]{E. Angelico}
\author[uc]{E. Spieglan}

\address[mymainaddress]{Incom Inc., 294 Southbridge rd, Charlton MA 01507 USA}
\address[anl]{Argonne National Laboratory, 9700 Cass Avenue, Lemont, IL, 60439 USA} 
\address[ssl]{Space Sciences Laboratory at University of California, 7 Gauss Way, Berkeley, CA 94720 USA}
\address[uc]{University of Chicago, 5640 S Ellis Ave, Chicago IL, 60637 USA}

\begin{abstract}
 We report on performance results achieved for recently produced LAPPDs - largest comercially available planar geometry photodetectors based on microchannel plates. These results include electron gains of up to $10^{7}$, low dark noise rates ($\sim$100 Hz/cm$^{2}$ at a gain of  $6\cdot10^6$), single photoelectron (PE) timing resolution of $\sim$50 picoseconds RMS (electronics limited), and single photoelectron spatial resolution along and across strips of 3.2mm (electronics limited) and 0.8 mm RMS respectively and high (about 25\% or higher in some units) QE uniform bi-alkali photocathodes. LAPPDs is a good candidate to be employed in neutrino experiments (e.g. ANNIE \cite {ANNIE}, WATCHMAN \cite{WATCHMAN}, DUNE \cite{DUNE}), particle collider experiments (e.g. EIC \cite{EIC}), neutrinoless double-beta decay experiments (e.g. THEIA \cite{THEIA}), medical and nuclear non-proliferation applications. 

\end{abstract}


\end{frontmatter}


\section{Introduction}

The Large Area Picosecond Photo-Detector (LAPPD\textsuperscript{TM}) is a microchannel plate (MCP) based planar geometry photodetector featuring single-photon sensitivity, semitransparent bi-alkali photocathode, millimeter spatial and picosecond temporal resolutions and an active area of 350 square centimeters. The"baseline" LAPPD employs a borosilicate float glass hermetic package. Photoelectrons are amplified with a stacked chevron pair of "next generation" large area MCPs produced by applying resistive and emissive Atomic Layer Deposition (ALD) coatings to glass capillary array (GCA) substrates. Signals are collected on microstrip anodes applied to the bottom plate.

Since 2015 a number of early commissioning trials performed at Incom Inc. demonstrated the ability to successfully seal LAPPD , apply uniform high QE photocathodes over the full area of the window, to achieve high gain from the chevron pair of ALD-GCA-MCPs, and to demonstrate single photoelectron sensitivity with saturated pulse height distributions \cite{CRAVEN2018}. These early trials culminated in the fall of 2017 with the fabrication of tiles that achieved all of these parameters at usable levels \cite{MINOT2018}.  In the late 2018 Incom demonstrated a capability of pilot production of LAPPDs with desired characteristics. 

Second generation LAPPDs are also currently under active development by Incom Inc., with our university collaborators \cite{HENRY2016}. Most notable is a next generation design that incorporates an anode capacitively coupled through a thin metal film deposited onto the inside bottom of the detector, to an application specific printed circuit board, positioned beneath the detector tile, outside of the vacuum package. This innovation will allow LAPPD to be customized for either stripline or pad readout, simplify connectivity, and facilitate use of LAPPD in high fluence applications. 

We report on performance characteristics of recently manufacured LAPPDs and discuss some of the features of second generation LAPPDs with capasitively coupled readout. 

\section{LAPPD design}

\begin{figure}[!h]%
\begin{center}%
\subfiguretopcaptrue
\subfigure[][] 
{
    \label{fig:LAPPD:design}
    \includegraphics[width=3.5cm]{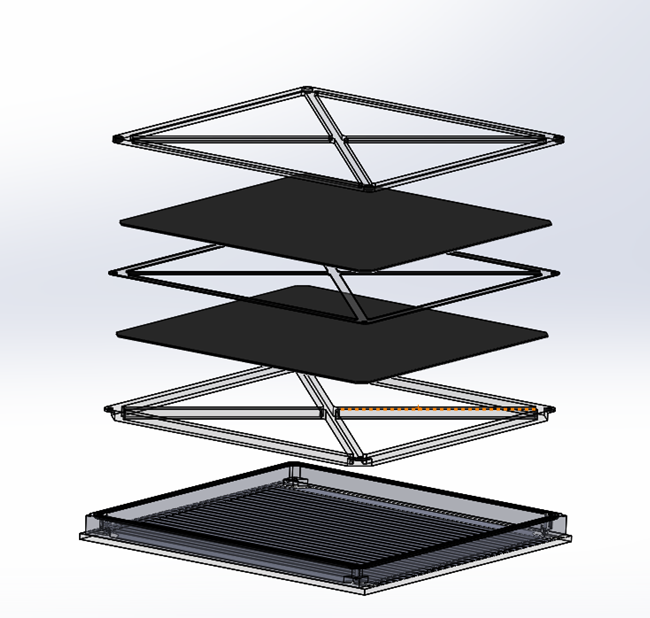}
} 
\subfigure[][] 
{
    \label{fig:LAPPD:photo}
    \includegraphics[width=4cm]{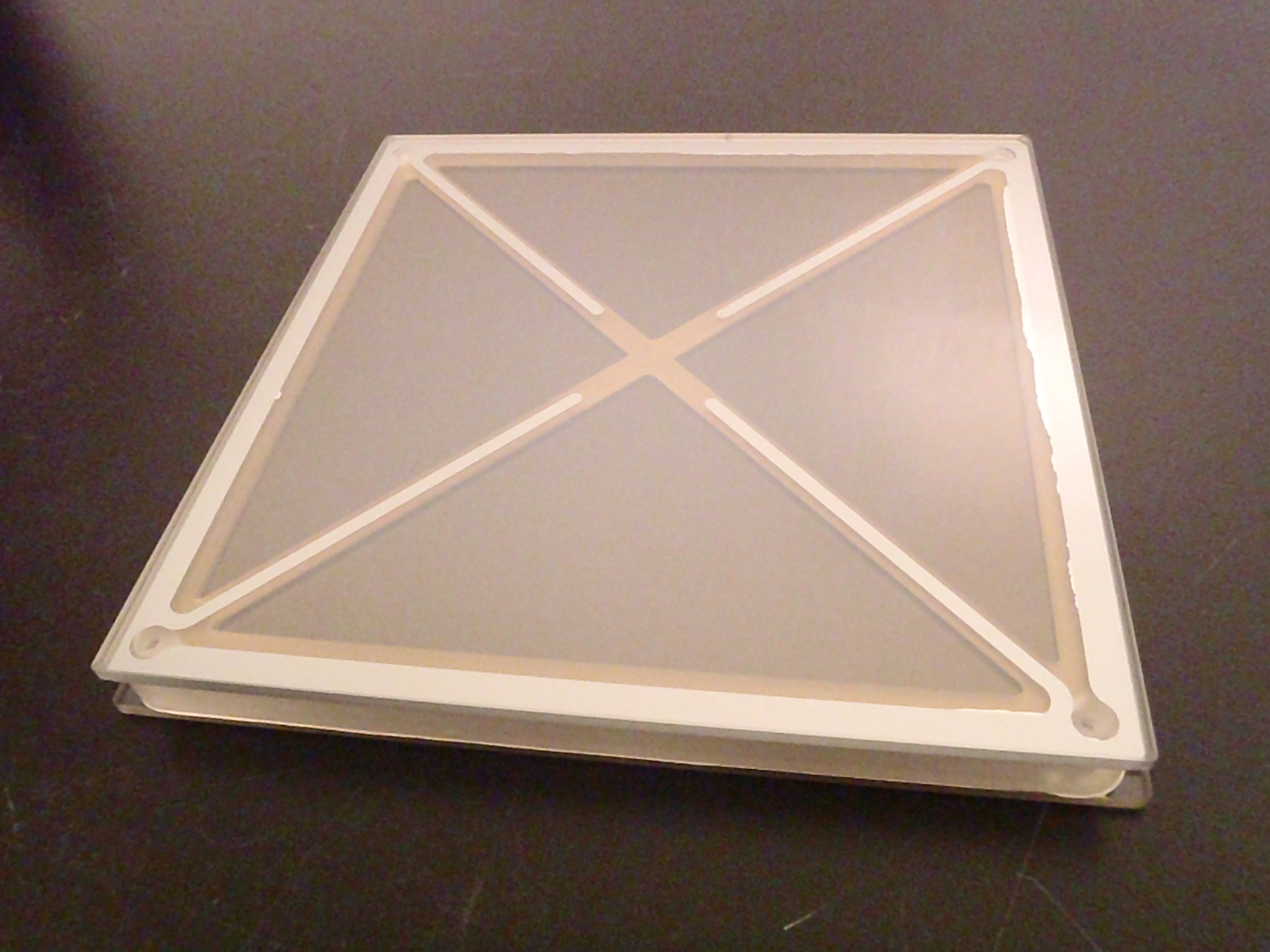}
} \caption{a) Exploded view of LAPPD showing glass detector enclosure, glass x-spacers and MCPs b) A photograph of an LAPPD. }
\label{fig:LAPPD}
\end{center}
\end{figure}

General design features and parameters for LAPPD incude borosilicate glass enclosure with no penetrating pins, a chevron pair of ALD functionalized MCPs, borosilicate glass spacers for rigidity, a choice of borosilicate or fused silica glass front window equipped with enhanced sensitivity Na$_2$KSb photocathode with active area of 350 cm$^2$, a microstrip anode with 28 silver strips for signal readout and independent biasing of photocathode and MCPs.   A schematic drawing of an exploded view of the LAPPD, with MCPs, X-spacer, and anode data strips is shown in \ref{fig:LAPPD:design}. A photograph of an LAPPD can be seen in \ref{fig:LAPPD:photo}.

\section{Experimental methods}

\begin{figure}
  \begin{center}
    \includegraphics[width=3cm]{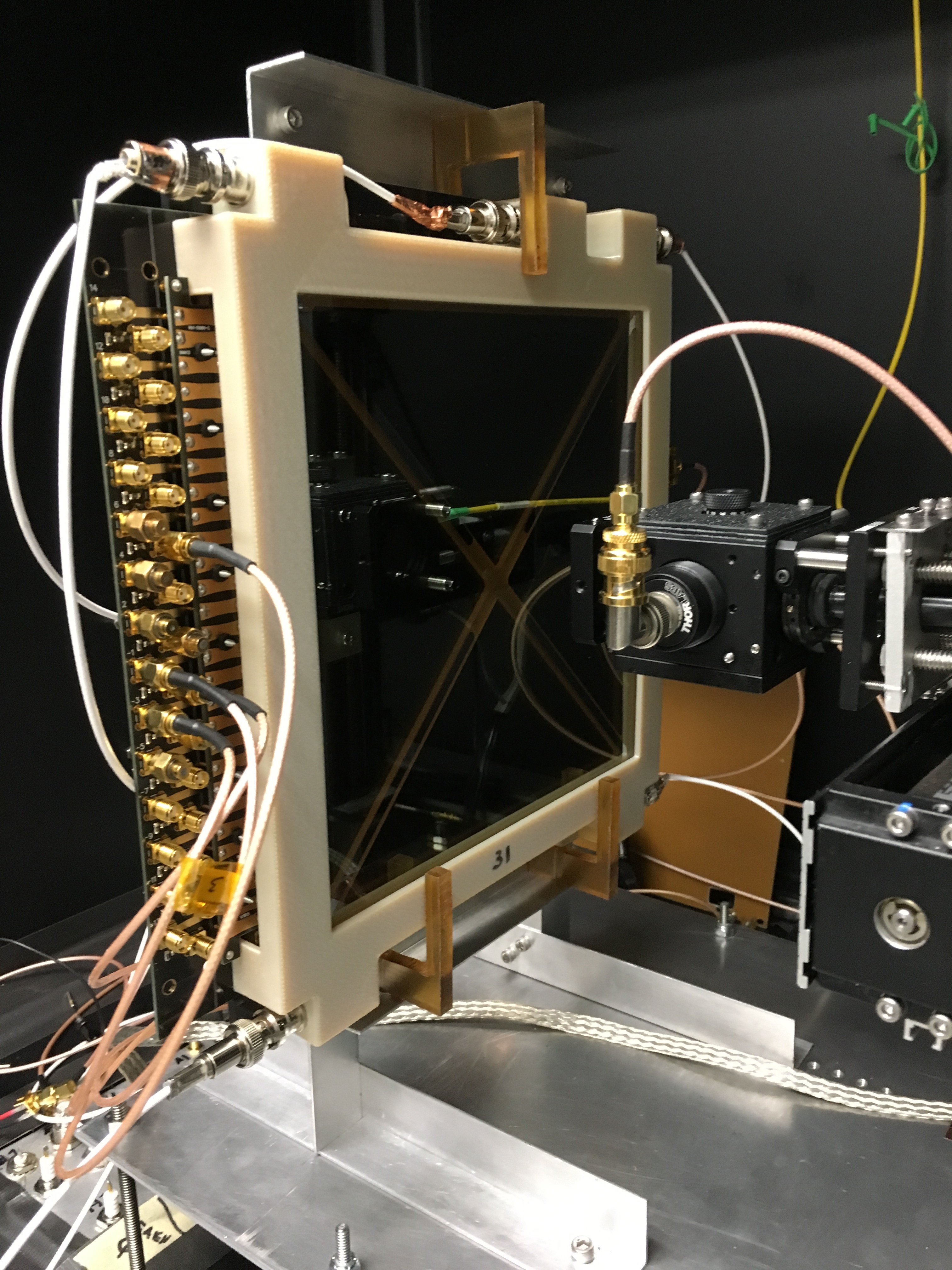}
    \caption{The LAPPD in the dark box enclosed in Ultem housing with high voltage connectors, and SMA connectors for signal readout.}
    \label{fig:test:station}
  \end{center}
\end{figure}

LAPPD performance tests were performed in a dark box with a UV light source and signal acquisition hardware. LAPPDs are provided with an Ultem housing that provides high voltage connections, and a backplane as shown in \ref{fig:test:station} that connects anode strips to SMA connectors with near-50 ohm impedance.  Gain was measured as a function of MCP voltage, using a charge sensitive amplifier and an ADC, and subsequently measured using DRS4 \cite{Ritt2008DesignAP} waveform samplers. 

\paragraph{Photocathode uniformity} The quantum efficiency of the photocathode was measured across the LAPPD window by scanning a 365 nm UV LED in an X-Y pattern of 3mm steps. The LED light was focused onto LAPPD window with a small lens resulting in an illuminated area of  $\sim$2.5 mm in diameter. The intensity of the input light was measured with a Thorlabs SM1PD2A photodiode, and a Keithley 6485 picoammeter. The photocurrent was collected and measured by connecting both sides of the entry MCP to a Keithley 2400 sourcemeter, with a 42 volt bias voltage between the MCP and the photocathode. The quantum efficiency is calculated from the ratio of these two quantities, less the dark current in each.
\paragraph{Gain measurements}  Single photoelectron MCP pulses for the gain measurement were produced by directing a 405 nm 63 ps FWHM laser pulse from PiLas model PiL040-FC laser to a selected point on the LAPPD window.  The laser was triggered externally at various rates. The trigger pulse was also used to provide a 12 $\mu$S window for the ADC, so the pulse height analyzer could detect charge pulses from the LAPPD response from the laser, if there were any, with minimal inclusion of dark pulses.  A neutral density filter (NE540B from Thorlabs) was used on the laser to reduce the intensity to the single photon level. Assuming the number of photoelectrons follows Poisson distribution the average number of photoelectrons per laser pulse should not exceed 0.22 \cite{Wang:2016xnu}. In this case the probability of producing more than one photoelectron is statistically suppressed at 90\% level. The LAPPD responded to 4 out of every 20 laser pulses thus satisfying this condition. Gain measurements were also made using waveform sampling where the time/charge area of each MCP pulse is used to infer the charge in the pulse.
\paragraph{Dark Count Rate} The Dark Count Rate was measured by a direct readout of LAPPD dark pulses with a 1GHz bandwidth oscilloscope at a threshold of 4mV. 
\paragraph{Timing Resolution} The time variation between the initiation of a photoelectron and the arrival of the MCP pulse at the end of a strip is of interest for timing applications. This variation represents the timing uncertainty of the LAPPD. The time variation was measured as follows. 63 ps FWHM 405 nm laser pulse was simultaneously fed onto a fast photodiode and tested LAPPD using a beam splitter. The laser intensity to LAPPD was reduced by neutral density filters to produce single photoelectrons. The time difference between the monitor pulse and the corresponding pulse from a single strip was measured by analyzing waveforms from DRS4 waveform samplers. There is also a $\sim$25 ps jitter in the width of the DRS4 timesteps, which is not corrected here. Hence, the quality of the measurement is somewhat environment-dependent, and the result here may not be the best achievable with the LAPPD. The LAPPD transit time variation may be extracted as a sum of squared variations as $\sigma_{meas}^{2}=\sigma_{LAPPD}^{2}+\sigma_{laser}^{2}$, where $\sigma_{meas}$ being the measured time variation, $\sigma_{LAPPD}$ is the Transit Time Spread of the tested LAPPD and $\sigma_{laser}$ is the width of the laser pulse.
\paragraph{Spatial Resolution} DRS4 waveform samplers were also used to determine spatial resolutions for single photoelectrons scanning both along-strips as well as across-strips. In the presented version of LAPPD the strips were 5.2 mm wide spaced by 1.7 mm gaps. X-Y position of the MCP charge cloud deposition may be measured with the LAPPD microstrip anode as follows. Along a strip, the position of the charge pulse may be inferred by measuring the relative time of arrival of pulses at each end of the strip, as the charge deposited by the MCP makes its way to ground at both ends. Timing variability at a given position provides the uncertainty of the position. Across-strip position is determined from a "center of mass" calculation that uses charge measured on five (or more) adjacent strips. The centroids are derived using distribution of the signal between five (or more) adjacent as a function of incident laser cross-strip position.  

\section{Results} 

\begin{figure}[h]%
\begin{center}%
\subfiguretopcaptrue
\subfigure 
{
    \label{fig:QE:37}
    \includegraphics[width=4.1cm]{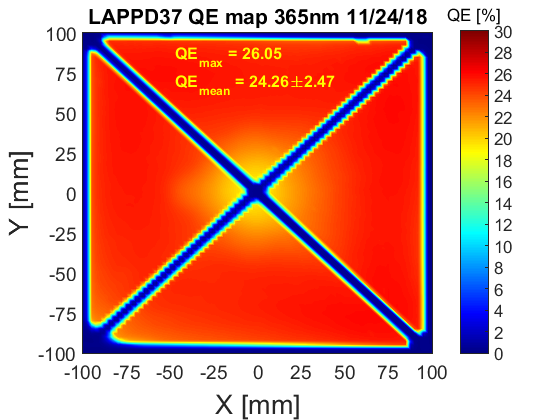}
} \hspace{0cm}
\subfigure 
{
    \label{fig:QE:38}
    \includegraphics[width=4.1cm]{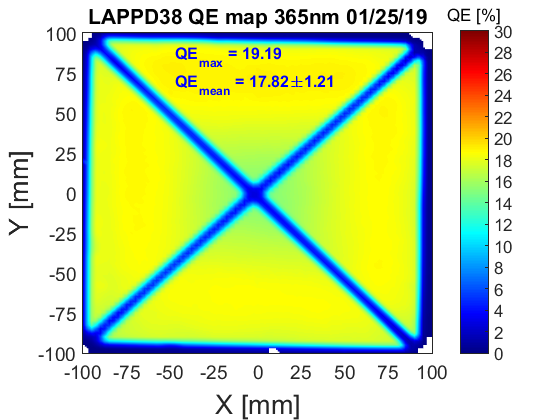}
}
\subfigure 
{
    \label{fig:QE:39}
    \includegraphics[width=4.1cm]{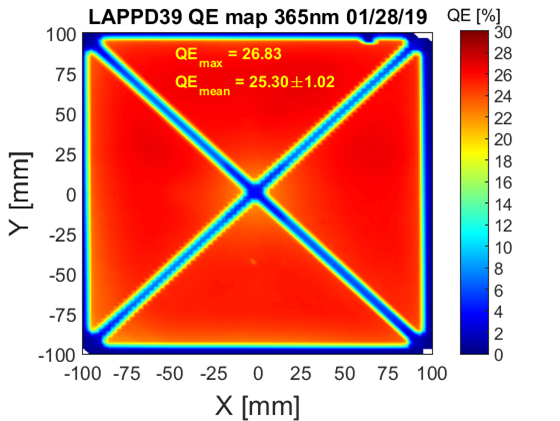}
}
\subfigure 
{
    \label{fig:QE:40}
    \includegraphics[width=4.1cm]{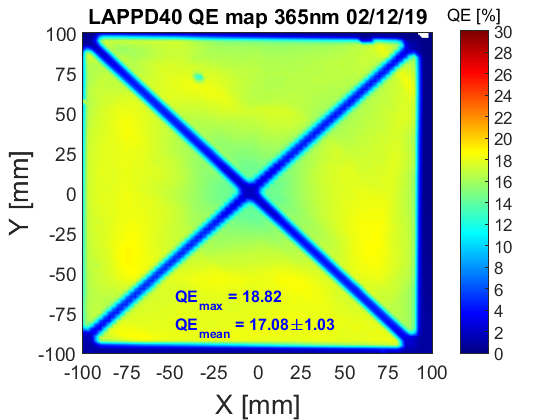}
}
\caption{QE maps measured for LAPPD37, LAPPD38, LAPPD39 and LAPPD40. Average QE values are indicated in the figures.}
\label{fig:QE}
\end{center}
\end{figure}

\paragraph{Photocathode quality} Examples of QE scans for the four consecutively sealed LAPPD37, LAPPD38, LAPPD39 and LAPPD40 measured at 365 nm are shown in \ref{fig:QE}. The mean QE at 365 nm was measured to be 24.3\%, 17.8\%,  25.3\%  and 17.1\% respectively. Lower QE measured in LAPPD38 and LAPPD40 are due to non-optimal photocathode deposition (temperature, vacuum etc.). At optimal deposition conditions QE values exceeding $>$20\% were demonstrated.     

\begin{figure}[h]%
\begin{center}%
\subfiguretopcaptrue
\subfigure 
{
    \label{fig:PHD:39}
    \includegraphics[width=4.1cm]{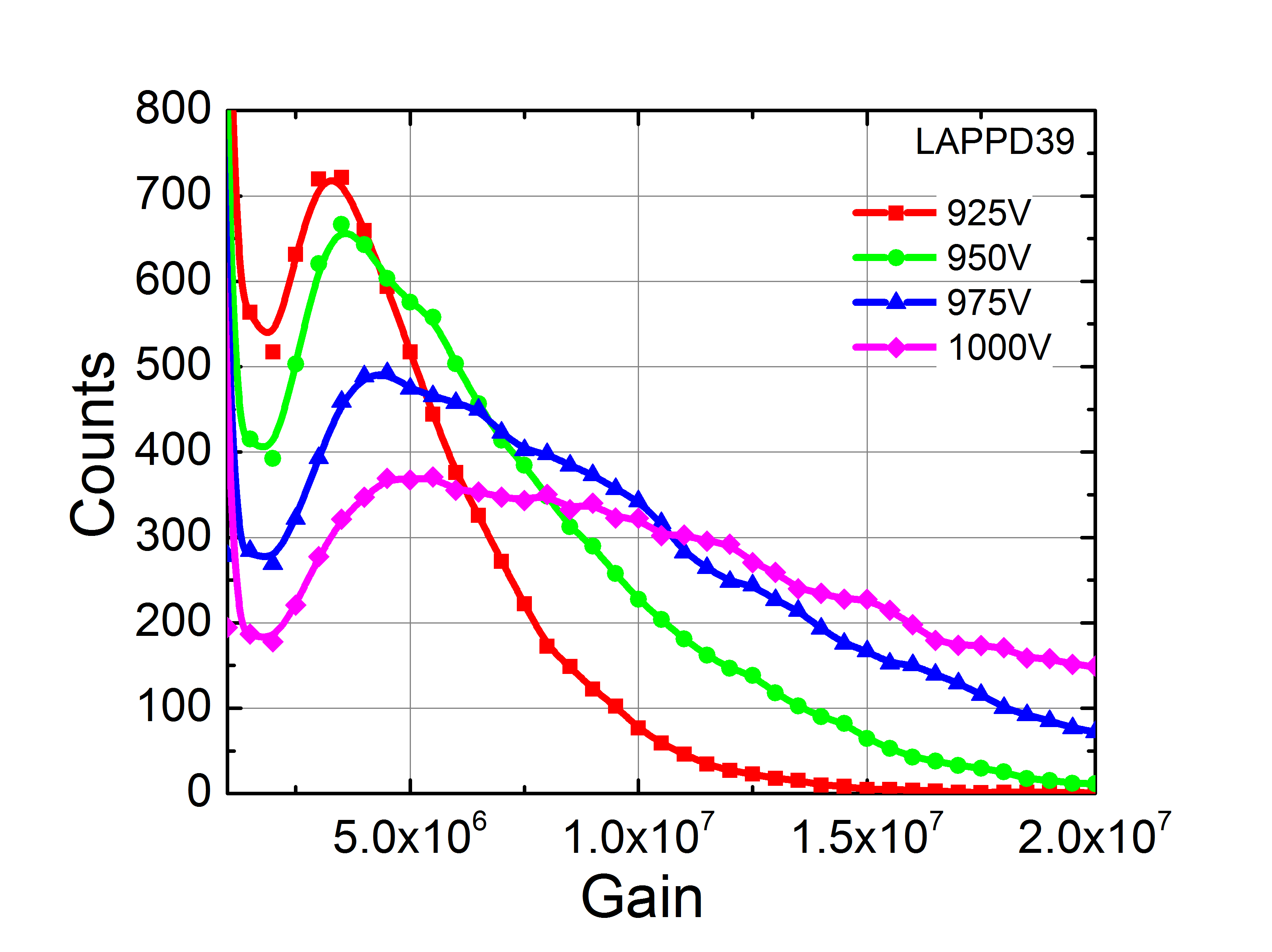}
} \hspace{0cm}
\subfigure 
{
    \label{fig:PHD:40}
    \includegraphics[width=4.1cm]{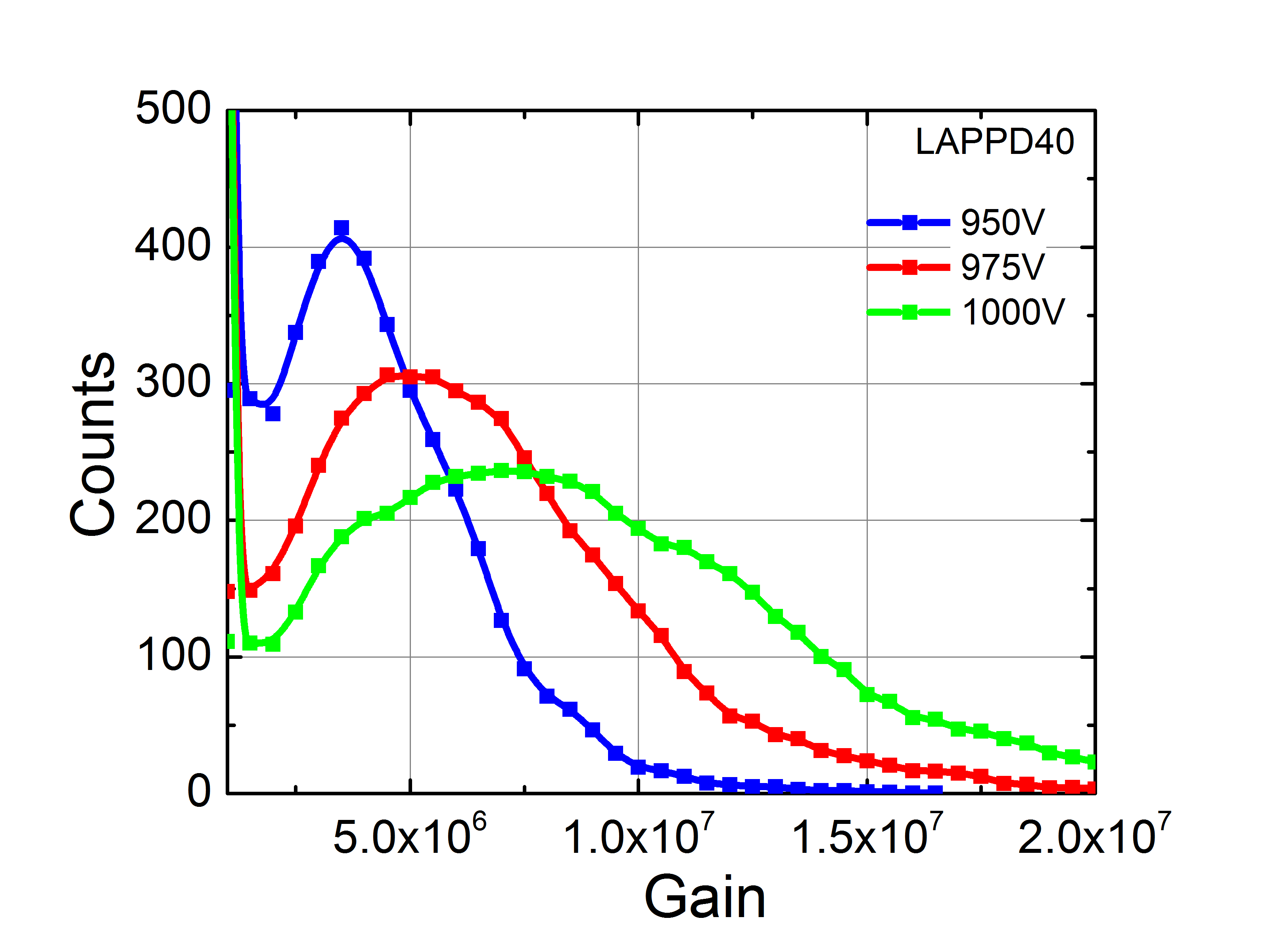}
}
\subfigure 
{
    \label{fig:G:39}
    \includegraphics[width=4.1cm]{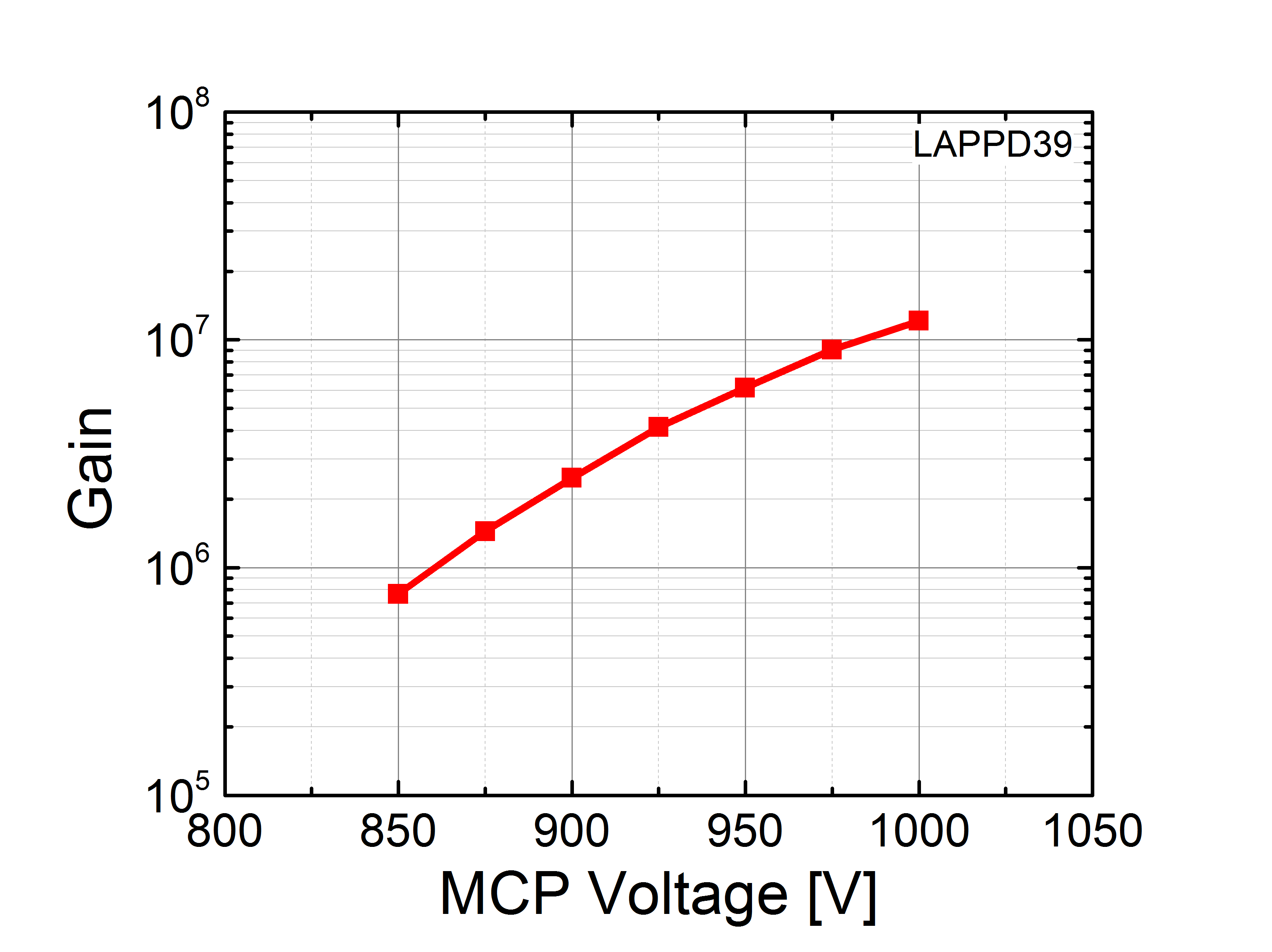}
}
\subfigure 
{
    \label{fig:G:40}
    \includegraphics[width=4.1cm]{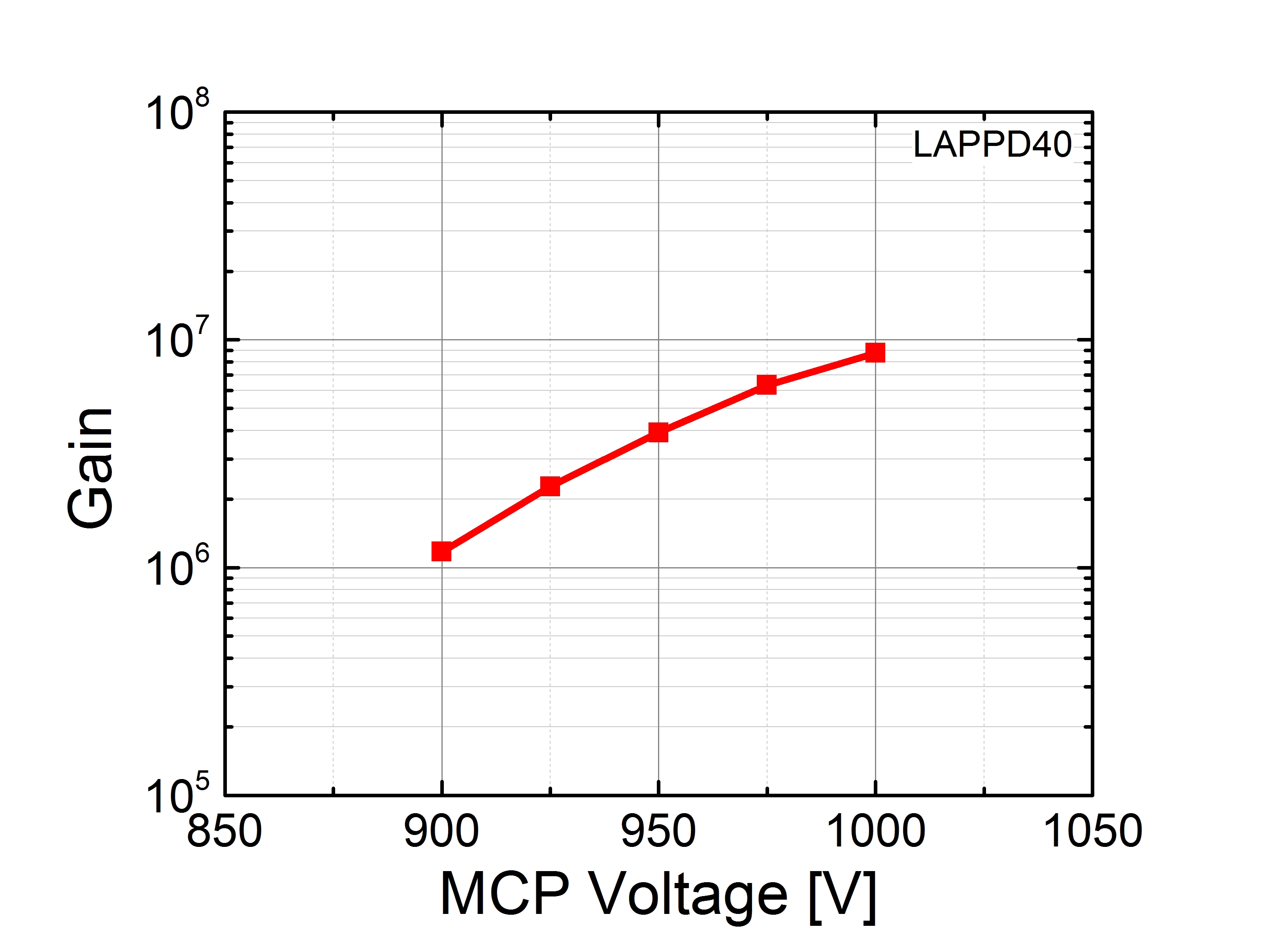}
}
\caption{Single photoelectron pulse height distributions measured in LAPPD39 and LAPPD40 at various MCP voltages. The respective gain values calculated from the pulse height distributions are also shown.}
\label{fig:gain}
\end{center}
\end{figure}

\paragraph{Pulse height distributions and gain} At a high gain the MCPs operate in saturation mode in which the electron-avalanche size is somewhat confined due to space charge effects. This leads to a peaked pulse height distribution for single photoelectrons. Detector gain at a given threshold can then be calculated as the average of the single photoelectron pulse height distribution. Examples of the single photoelectron pulse height distributions and corresponding gains measured in LAPPD39 and LAPPD40 are presented in \ref{fig:gain}. To plot the pulse height distributions shown in \ref{fig:gain} a data set of 10000 waveforms recorded with DRS4 digitizers was analyzed.  In the recent LAPPDs a gain of 10$^7$ was achieved at a reasonably low dark count rate (see below) and MCP voltage setting. 

\begin{figure}[h]%
\begin{center}%
\subfiguretopcaptrue
\subfigure 
{
    \label{fig:dark:39}
    \includegraphics[width=4cm]{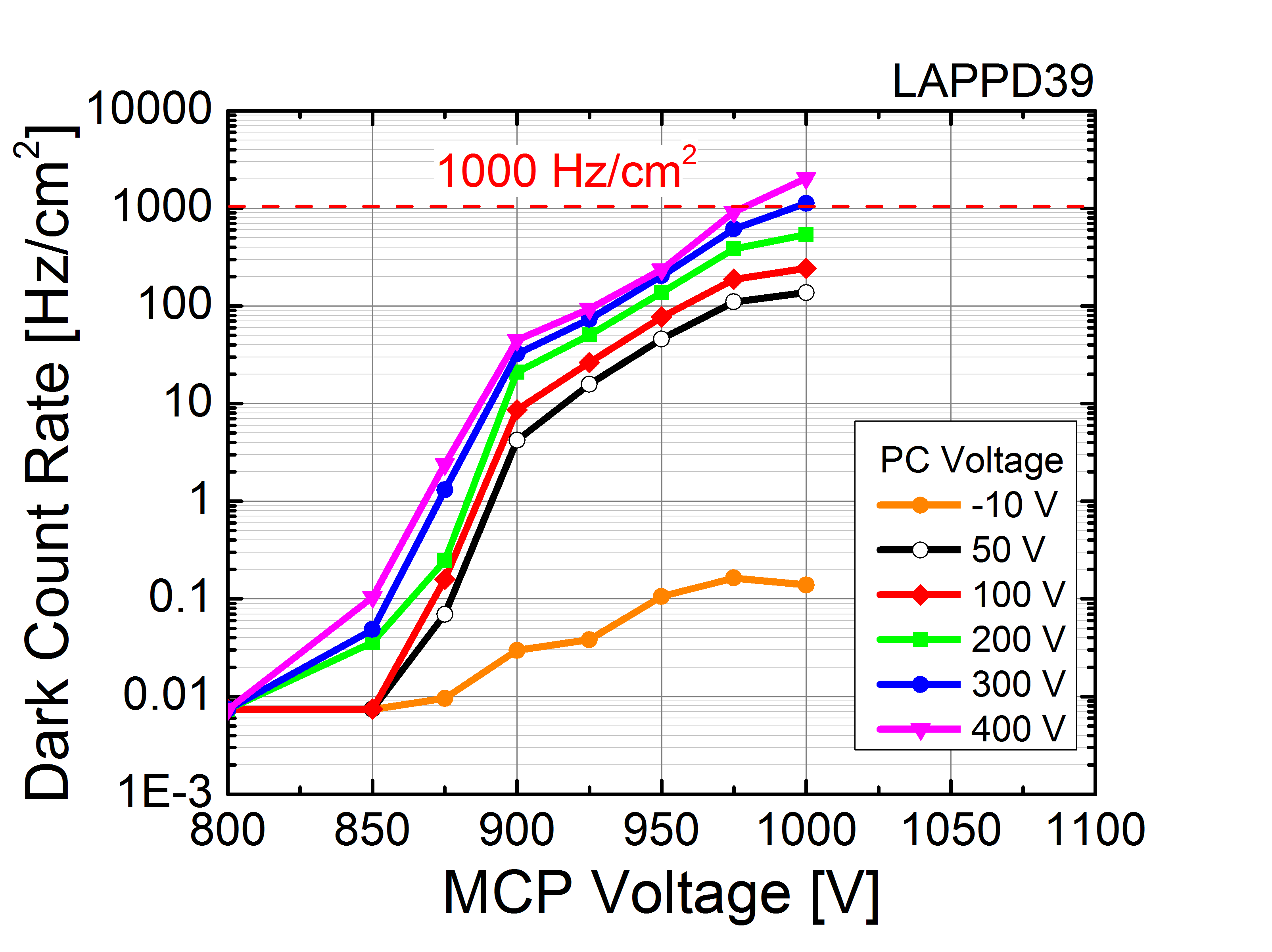}
} 
\subfigure
{
    \label{fig:dark:40}
    \includegraphics[width=4cm]{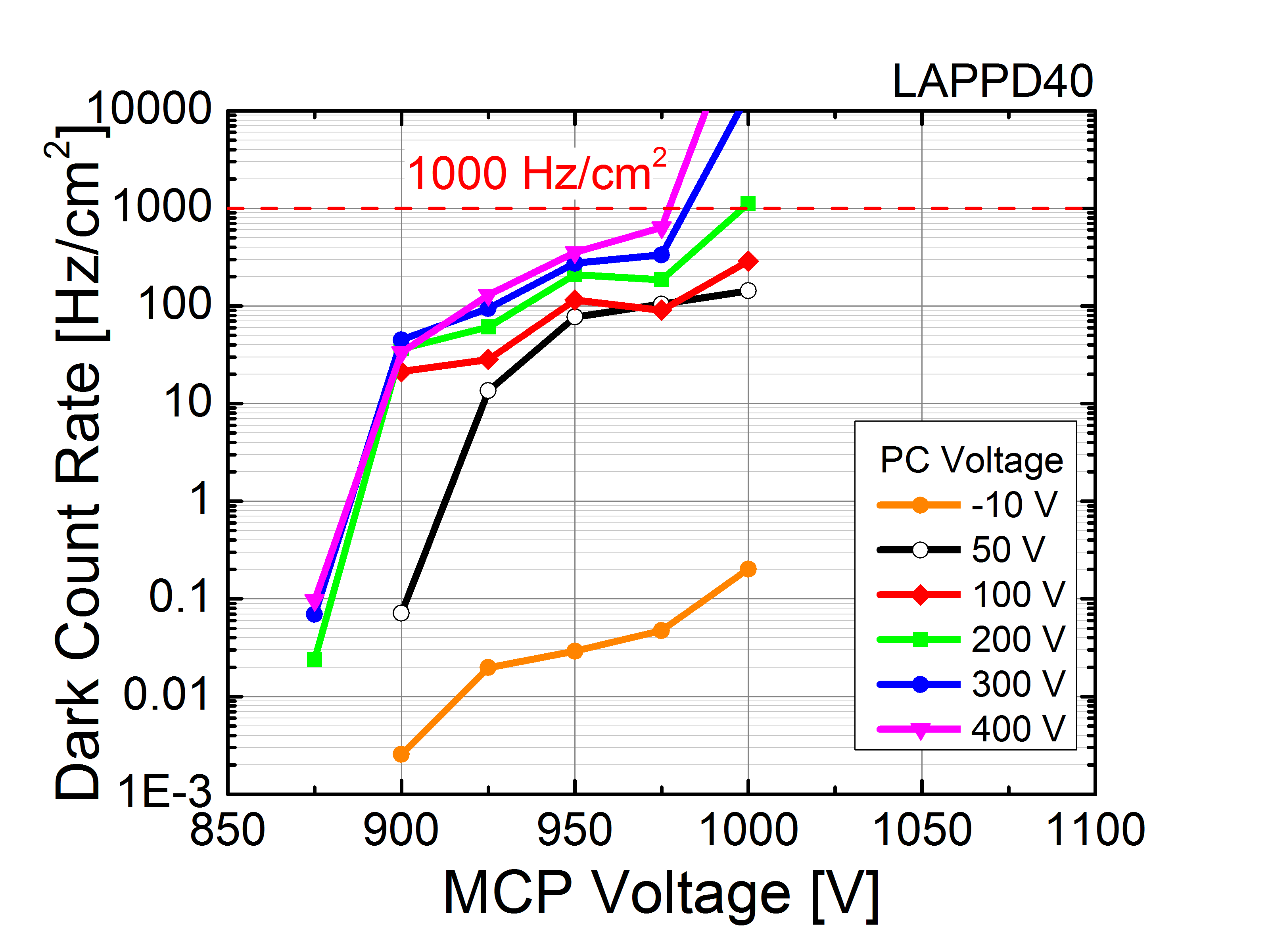}
} \caption{Dark count rate as a function of MCP voltage measured at various photoelectron extraction fields or photocathode voltages. }
\label{fig:dark}
\end{center}
\end{figure}

\paragraph{Dark count rate} The dark count rate has been tremendously reduced compared to earlier tiles as shown in \ref{fig:dark} on example of LAPPD39 and LAPPD40. In LAPPD39 the dark rate stayed below 1000 Hz/cm$^2$ (dashed red line in the plots) within the full range of MCP and photocathode voltages up to MCP gain of $10^7$. At optimal operation conditions with 200V at the photocathode and a gain of $6\cdot10^6$ the observed dark count rate was only 100 Hz/cm$^2$. For LAPPD40 the dark count rate was well below 1000 Hz/cm$^2$ up to a gain of $6\cdot10^6$. The orange line on the plots represents the dark rate by MCPs only with the reversed photoelectron extraction field. 

\begin{figure}[h]%
\begin{center}%
\subfiguretopcaptrue
\subfigure[][] 
{
    \label{fig:TTS:40}
    \includegraphics[width=4.2cm]{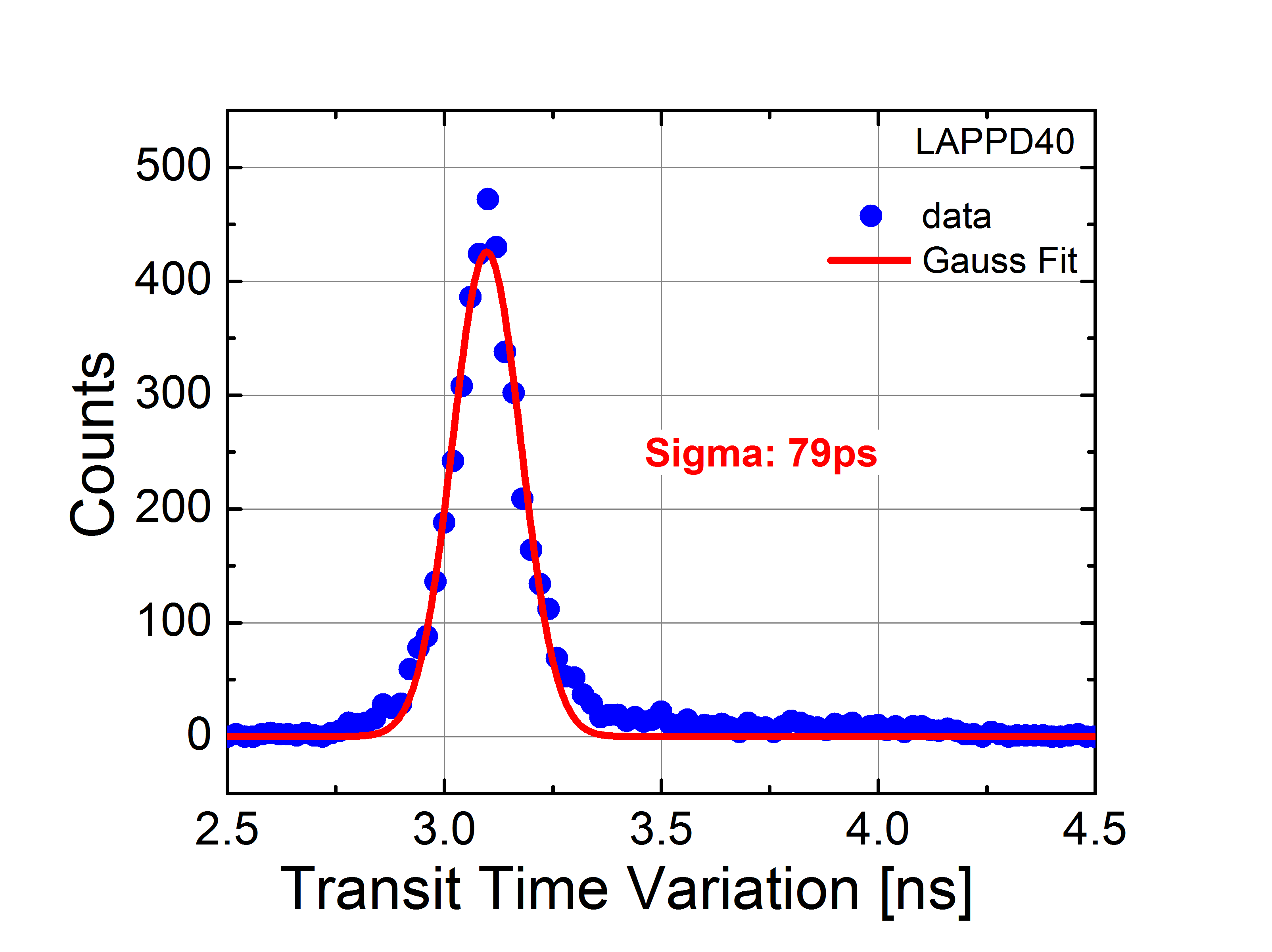}
} 
\subfigure[][] 
{
    \label{fig:TTSPC}
    \includegraphics[width=4cm]{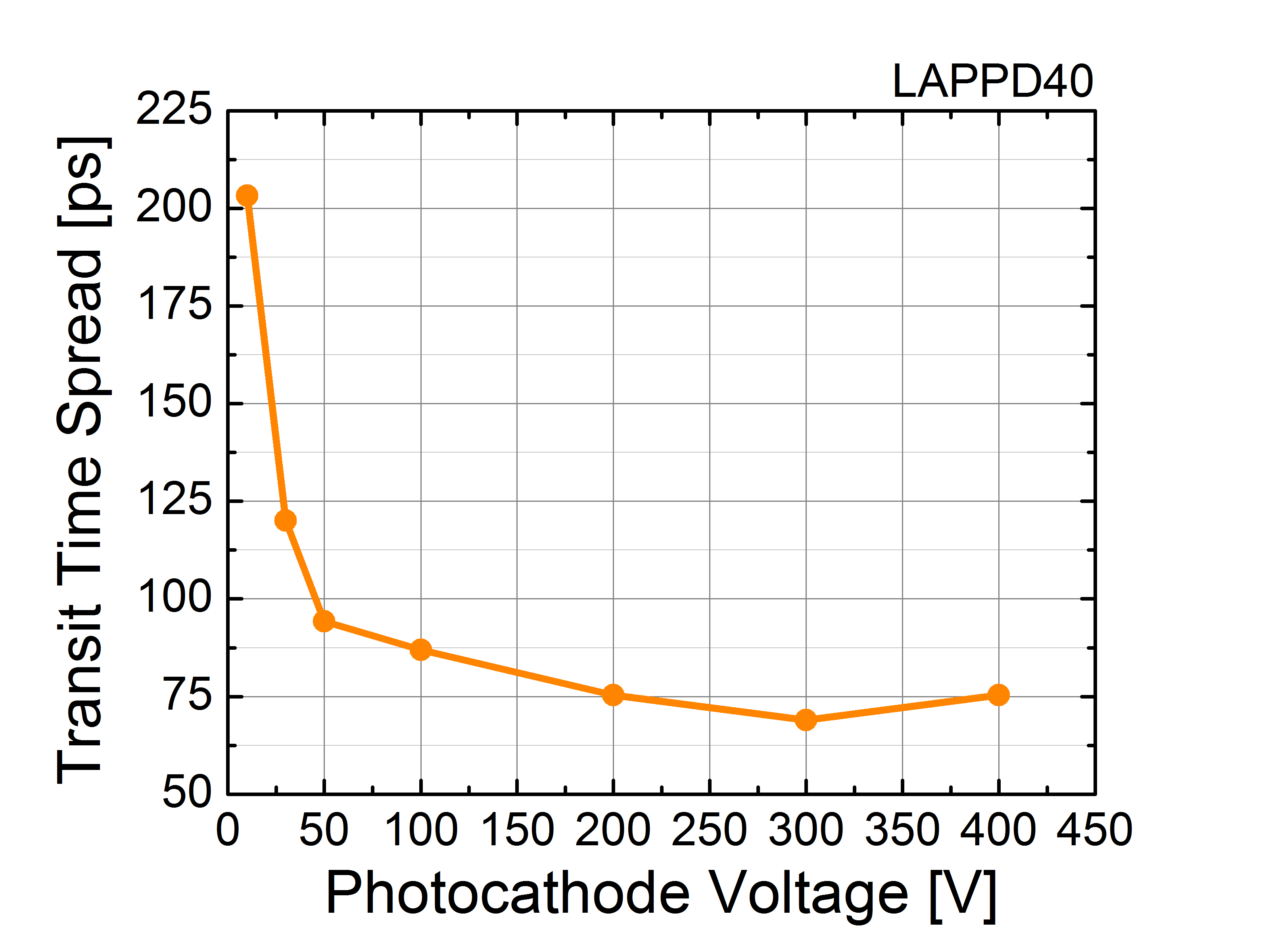}
} \caption{a) Transit Time Variation measured in LAPPD40 at 400V bias between the photocathode and top face of the entry MCP; b) Transit Time Variation as a function of photoelectron extraction field.}
\label{fig:TTS}
\end{center}
\end{figure}

\paragraph{Timing Resolution} A key feature of LAPPD is its picosecond timing ability. Arrival time jitter of photoelectron signal measured in LAPPD40 at 400V bias between the photocathode and top face of the entry MCP is shown in \ref{fig:TTS:40}. From a Gaussian fit the measured Transit Time Variation $\sigma_{meas}$ was calculated to be 79 ps. Given the assumption above that $\sigma_{meas}^{2}=\sigma_{LAPPD}^{2}+\sigma_{laser}^{2}$, the LAPPD transit time variation $\sigma_{LAPPD}$ was calculated to be about 50 ps. It was also shown in \ref{fig:TTSPC} that the measured transit time variation is a function of photoelectron extraction field. At a photoelectron extraction field that corresponds to 200V potential difference between the photocathode and the entry MCP $\sigma_{meas}$ approaches about 80 ps (that corresponds to $\sigma_{LAPPD}$ of 50 ps). 

\begin{figure}[h]%
\begin{center}%
\subfiguretopcaptrue
\subfigure[][] 
{
    \label{fig:XPOS:DIST:31}
    \includegraphics[width=3.8cm]{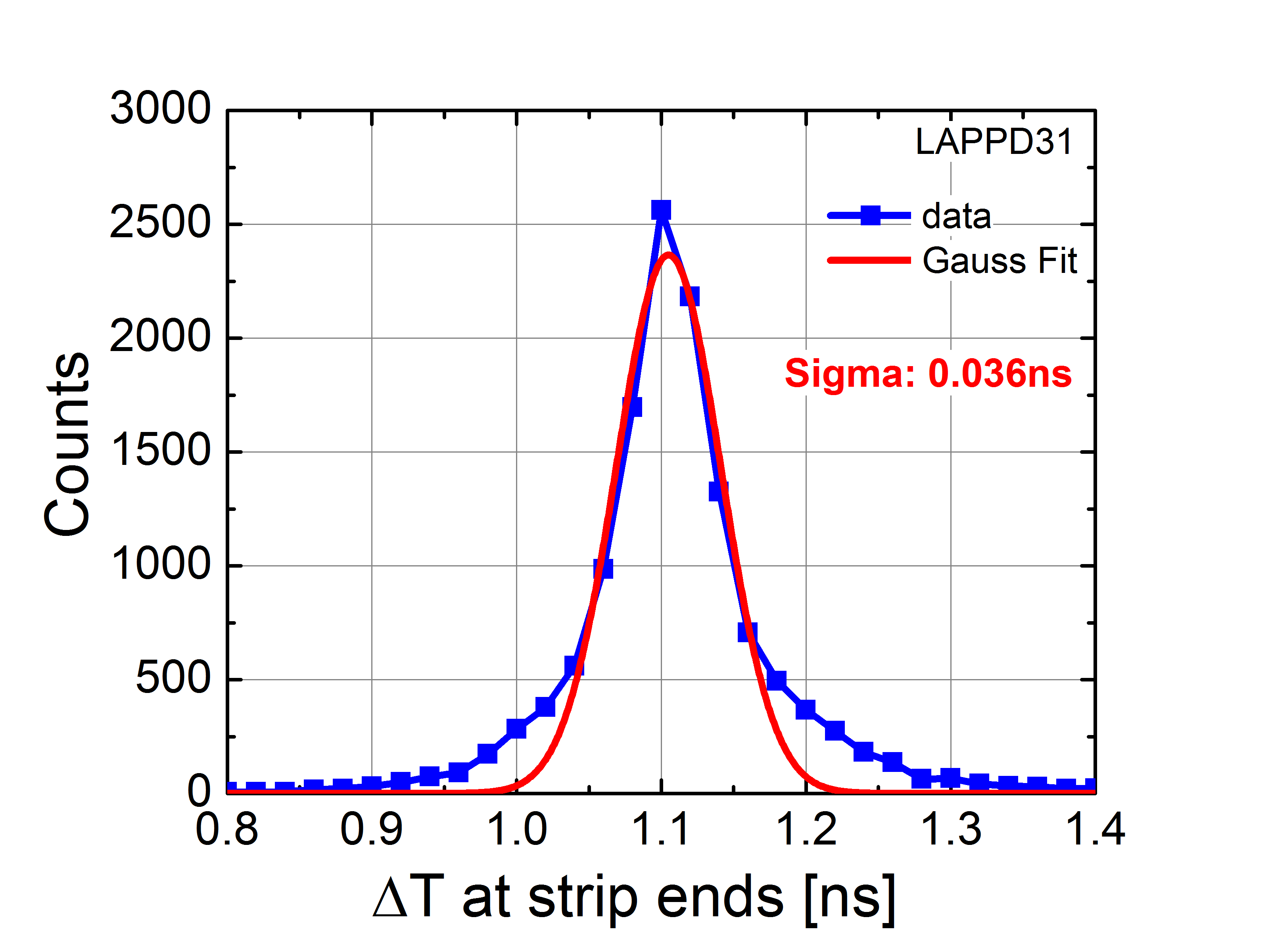}
} \hspace{0cm}
\subfigure[][] 
{
    \label{fig:XPOS:LAS:31}
    \includegraphics[width=4.5cm]{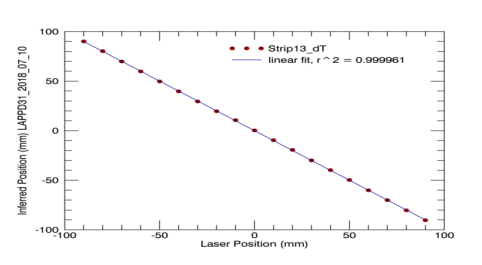}
}
\subfigure[][] 
{
    \label{fig:YPOS:DIST:31}
    \includegraphics[width=3cm]{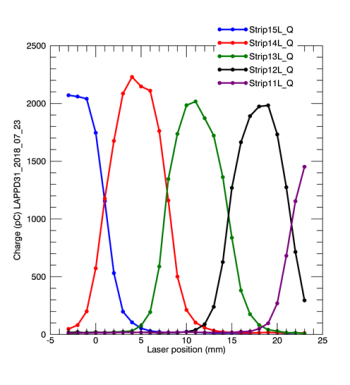}
}
\subfigure[][] 
{
    \label{fig:YPOS:REC:31}
    \includegraphics[width=4.3cm]{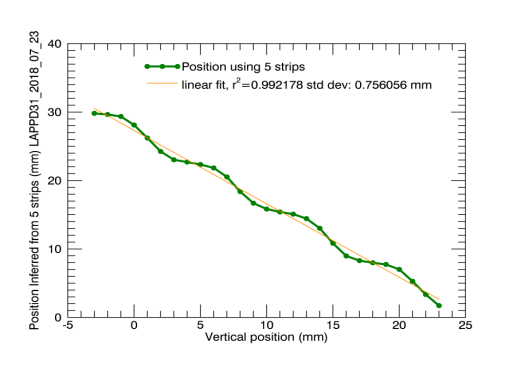}
}
\caption{Measurements of spatial resolution of LAPPD (example of LAPPD31). a) Distribution of relative arrival times at a fixed laser position on the strip. Position resolution can be calculated from a Gaussian fit. b) Reconstructed position along the strip as a function of laser position. c) Distribution of the signal amplitude as a function of the laser position measured in 5 adjacent strips. d) Reconstructed position across the strips as a function of laser position. Standard deviation from the linear fit defines the position resolution.}
\label{fig:POS}
\end{center}
\end{figure}

\paragraph{Spatial Resolution} In LAPPD31 the relative arrival time sigma of pulses observed at both ends of a strip was 36 ps at one position, as shown in \ref{fig:XPOS:DIST:31}. Signal transmission speed along the strip was determined by measuring differential time between two ends of a strip varying the position of the laser spot along the strip  \cite{Wang:2016xnu}. At a scale of 11.4 ps/mm this leads to a position uncertainty of 3.2 mm. This result does not include 25 ps uncertainty in the DRS4 timestamps so it does not fully represent the best achievable resolution of LAPPD. Reconstructed signal position along the strip as a function of laser position is shown in \ref{fig:XPOS:LAS:31}.  Across-strip position determined from a centroid calculation that uses charge measured on five adjacent strips can be seen in \ref{fig:YPOS:DIST:31}. The centroids were derived using distribution of the signal between the five adjacent strip as a function of incident laser cross-strip position. The reconstructed position calculated from center of mass position in five adjacent strips is shown in \ref{fig:YPOS:REC:31}. Across strip position resolution calculated as a standard deviation from the linear fit was 0.76 mm for a strip pitch of 6.1 mm.

\begin{figure}[h]%
\begin{center}%
\subfiguretopcaptrue
\subfigure 
{
    \label{fig:PHOTO:38}
    \includegraphics[width=5.0cm]{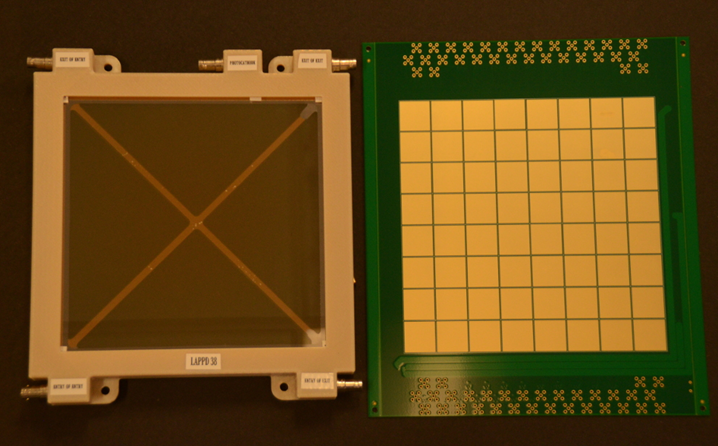}
} 
\subfigure 
{
    \label{fig:PHD:38}
    \includegraphics[width=3.5cm]{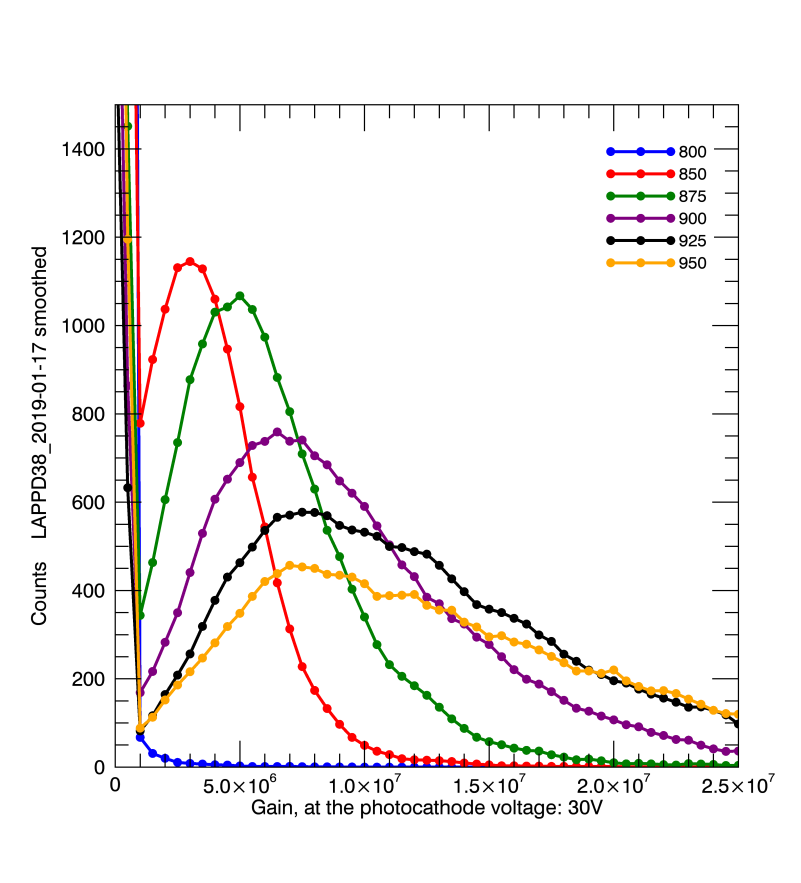}
}
\subfigure 
{
    \label{fig:G:38}
    \includegraphics[width=3.5cm]{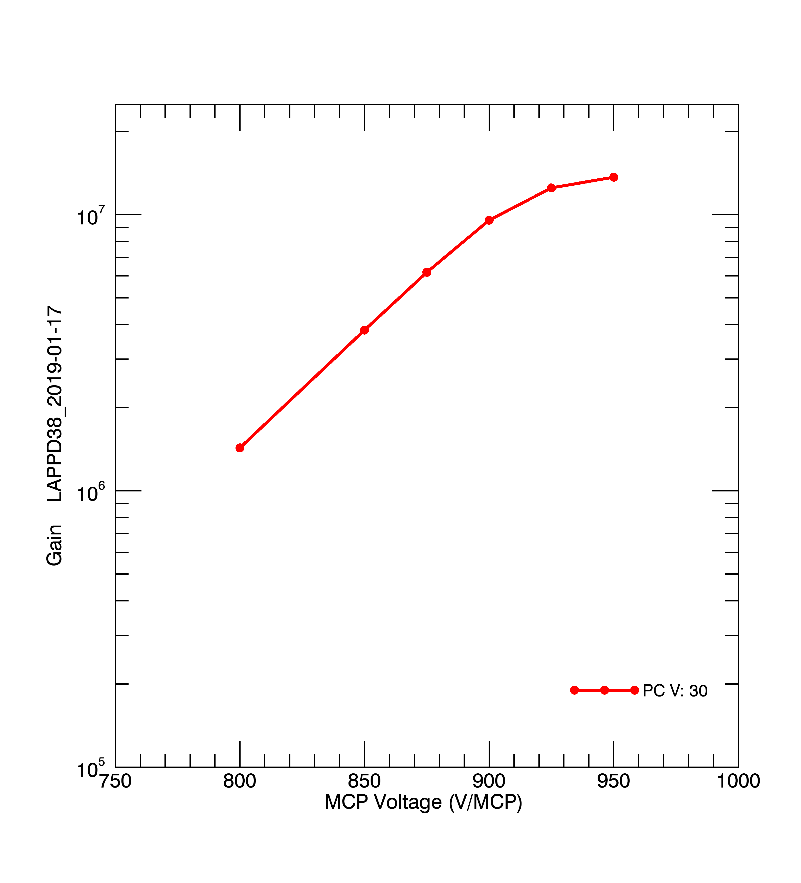}
}
\caption{Preliminary test results for the second generation LAPPD (LAPPD38). a) A photograph of sealed GEN II LAPPD in the Ultem housing. Pad readout PCB board is shown next to it. b) Single photoelectron pulse height distributions acquired at various MCP voltages. c) Gain as a function of MCP voltage derived from the distributions.}
\label{fig:GENII}
\end{center}
\end{figure}

\paragraph{GEN II LAPPD development}  As mentioned above second generation LAPPDs are characterized by an alumina ceramic body and capacitively coupled readout. At the time of writing several Gen II LAPPDs were successfully sealed. Some test results measured in LAPPD38 \ref{fig:PHOTO:38} are presented in \ref{fig:GENII}. A high QE photocathode was demonstrated in the ceramic package (\ref{fig:QE:38}. Single photoelectron pulse height distributions and gain as function of MCP voltage recorded directly from the anode are shown in \ref{fig:PHD:38} and \ref{fig:G:38} respectively. Another Gen II LAPPD LAPPD36 is being installed near the beamline in FNAL to investigate its performance in the experimental environment. 

\section{Summary}  Incom has brought LAPPD technology through a commissioning phase to a pilot production phase. Presently LAPPDs are being supplied to early adopters (e.g. ANNIE experiment)  for "real environment" testing. Recently produced LAPPDs exhibit the following characteristics:
\begin{itemize}
{
\item high gain of 10$^7$, 
\item high photocathode QE of up to 25\% ,
\item low noise of 100 Hz/cm$^2$ at a gain of $6\cdot10^6$, 
\item mm scale position resolution (electronics limited),
\item timing resolution of 50 ps (electronics limited).
}
\end{itemize}

These features make LAPPD a good candicate to be employed in a number of applications in neutrino and rare-decay experiments, particle collider experments and others. Several Gen II LAPPD tiles have been produced exhibiting similar performance as in Gen I LAPPDs. 

\section*{Acknowledgments}

This work supported by U.S. Department of Energy, USA, Office of Science, USA, Office of Basic Energy Sciences, USA, Offices of High
Energy Physics, USA and Nuclear Physics, USA under DOE contracts: DE-SC0009717; DE-SC0011262, and DE-SC0015267. We also acknowledge former Incom Inc. employee Christofer A. Craven for his significant contribution to the project.

\bibliography{mybibfile}

\end{document}